\documentclass[english]{article}
\usepackage[T1]{fontenc}
\usepackage[latin1]{inputenc}

\makeatletter

\let\SF@@footnote\footnote
\def\footnote{\ifx\protect\@typeset@protect
    \expandafter\SF@@footnote
  \else
    \expandafter\SF@gobble@opt
  \fi
}
\expandafter\def\csname SF@gobble@opt \endcsname{\@ifnextchar[
  \SF@gobble@twobracket
  \@gobble
}
\edef\SF@gobble@opt{\noexpand\protect
  \expandafter\noexpand\csname SF@gobble@opt \endcsname}
\def\SF@gobble@twobracket[#1]#2{}

 \newenvironment{lyxlist}[1]
   {\begin{list}{}
     {\settowidth{\labelwidth}{#1}
      \setlength{\leftmargin}{\labelwidth}
      \addtolength{\leftmargin}{\labelsep}
      }}
   {\end{list}}

\usepackage{babel}
\makeatother
\begin{document}

\title{Alchemistry of the P versus NP question }

\author{Bonifac Donat%
\footnote{Pr. Emer. Ecole Normale Supérieure Polytechnique - Antsiranana 201
- MADAGASCAR. bdonat@acm.org%
}}

\maketitle
\begin{abstract}
Are P and NP provably inseparable ? Take a look at some unorthodox,
guiltily mentioned folklore and related unpublished results. 
\end{abstract}

\section{P versus NP at the Speed of Light}

Intuitively, P == NP %
\footnote{Or P = NP.%
} looks like : 

\begin{itemize}
\item There is a polynomial machine%
\footnote{A polynomial time-bouded Turing machine.%
} M$_{\textrm{m}}$ that correctly guesses a satisfying line of truth-values
for every input x in Sat %
\footnote{Set of all Boolean expressions in CNF (Conjunctive Normal Form) which
can be satisfied by some truth-value line x , as an instance of Sat
is coded by an adequate binary string.%
}. 
\end{itemize}
More formally :

\begin{itemize}
\item there is a Turing machine M$_{\textrm{m}}$ of Gödel number m and
\item there are integers a, b > 0 so that, if x is in Sat and the output
M$_{\textrm{m}}$(x) is a satisfying line for x, then the number of
cycles of M$_{\textrm{m}}$ over x is \\
t$_{\textrm{m}}$(x) < |x|$^{\textrm{a}}$ + b. %
\footnote{Whose length is |x|, the number of bits of x.%
} %
\footnote{Translatable into any formulation of computability.%
}
\end{itemize}
Its negation P < NP %
\footnote{Or P =! NP, or P != NP.%
} can of course be expressed as an arithmetical sentence. %
\footnote{The real fun starts when writing down the formal versions for P ==
NP as $\Sigma$$^{\textrm{0}}$$_{\textrm{2}}$ and P =! NP as $\Pi$$^{\textrm{0}}$$_{\textrm{2}}$
sentences.%
}

\section{P < NP's Milky Way}

Suppose %
\footnote{After a too long nap during a talk by someone from the CS department.%
} you believe that P < NP. Then you should find a hard problem, or
a hard instance of some NP-complete problem. %
\footnote{So, there is no ``hard instance,'' but an infinite hard set, if P
< NP holds.%
}

\subsection{SAT's trivial Folklore }

Given any instance x in Sat, there is always some polynomial algorithm
that settles it. %
\footnote{Outputs a satisfiable line for it.%
}

\subsection{Poly-Algos' Folklore }

Let n be any natural number. Then there are infinitely many polynomial
algorithms that :

\begin{itemize}
\item Settle all instances of Sat up to n plus infinitely many other of
its instances.
\item If P$_{\textrm{i}}$ and P$_{\textrm{j}}$ are two of those polynomial
algorithms, then we can have that P$_{\textrm{j}}$ accepts strictly
more instances of Sat than P$_{\textrm{i}}$.%
\footnote{Proof at hand.%
}
\item Then recall that to prove P < NP you are supposed to find an infinite
, recursive family of Sat's instances, so that for each polynomial
machine P$_{\textrm{n}}$ there is an instance x$_{\textrm{n}}$ in
Sat for which P$_{\textrm{n}}$ fails to output a satisfying line
of truth values.%
\footnote{Those are recursively enumerable sets.%
}
\end{itemize}
If you need a sketch : 

\begin{itemize}
\item First, enumerate%
\footnote{Using the so-called '\cite{BGS} trick' : Every poly machine can be
written as a couple < M , C$_{\textrm{p}}$> %
} all polynomial algorithms. %
\footnote{M is aTuring machine, C$_{\textrm{p}}$ is a clock that shuts down
the operation of M overbinary input x after p(|x|) cycles, p is a
polynomial with positive integer coefficients (notice that every polymachine
can be written as one such couple).%
}
\item Then, for each polynomial algorithm in the sequence you must prove
that it fails to output a satisfying line at least once. %
\footnote{Notice that P =! NP implies : if some algorithm fails once, it will
fail infinitely many times.%
}
\item Then find some convention to deal with the situations where the clock
actually interrupts the operation of M. %
\footnote{And cumulate your prize with many others in various disciplines.%
}
\end{itemize}

\section{P == NP's Black Hole}

Suppose that %
\footnote{after drinking the good wine of hopeful expectations laced with some
twisted logic.%
} you definitely go against current wisdom, and think that P == NP.
That is to say, you think that there actually is a polynomial algorithm.

\subsection{PA%
\footnote{Peano Arithmetic.%
}'s Folklore }

In an informal guise : If P == NP is true, then you can prove it in
some version of arithmetic. %
\footnote{This means, we only require arithmetic tools to prove P == NP, if
it turns out to be true.%
} %
\footnote{Proof in detail can be found in \cite{B-DH}%
}

\begin{itemize}
\item First, the fact that P == NP holds of the standard integers implies
that P == NP can be written as a $\Pi$$^{\textrm{0}}$$_{\textrm{1}}$
sentence%
\footnote{For the definition of $\Pi$$^{\textrm{0}}$$_{\textrm{1}}$sentences,
see \cite{B-DH}%
}.
\item Then it is provable in PA$_{\textrm{1}}$ %
\footnote{PA + all true $\Pi$$^{\textrm{0}}$$_{\textrm{1}}$ sentences. %
}. %
\footnote{Either PA proves it, or PA + one extra-$\Pi$$^{\textrm{0}}$$_{\textrm{1}}$sentence.%
}
\end{itemize}

\subsection{PA$_{\textrm{1}}$'s Folklore }

If P < NP is independent of PA$_{\textrm{1}}$, then it holds true
of the standard integers.%
\footnote{And far beyond.%
}

\subsection{ZFC$_{\textrm{1}}$'s Folklore }

ZFC$_{\textrm{1}}$ is the system %
\footnote{Nonrecursive.%
} that consists of ZFC %
\footnote{Zermelo-Fraenkel + Axiom of Choice.%
} plus all arithmetic $\Pi$$^{\textrm{0}}$$_{\textrm{1}}$ sentences.

\begin{itemize}
\item Suppose that this theory has a model with standard arithmetic.
\item If P < NP is independent of ZFC$_{\textrm{1}}$, then it holds true
of the standard integers.
\end{itemize}

\subsection{ZFC's Folklore}

If P < NP is independent of ZFC %
\footnote{Which is again supposed to have a model with standard arithmetic.%
} then it holds true of the standard integers too. %
\footnote{Prove it first.%
} 

The local strategy becomes : 

\begin{itemize}
\item First show that ZFC %
\footnote{No primes or indices.%
} proves the following : P < NP if and only if R %
\footnote{R is a recursive function described in \cite{B-DH}%
} is total.
\item Then prove : for every polynomial machine m there is a Sat instance
x =< R(m) so that polynomial machine m fails to output a satisfying
line at input x. %
\footnote{Which is naively $\Pi$$^{\textrm{0}}$$_{\textrm{1}}$, and equivalent
to P =! NP. %
} %
\footnote{x =< R(m) means that instance x is bounded by R(m).%
} %
\footnote{And find a title for the speech at Claymaths'.%
}
\end{itemize}

\section{The Busy Beaver creeps up to the Stars}

One can alternatively think of the following (quasi)-algorithm for
Sat :

\begin{enumerate}
\item List the BGS machines in the usual way : P$_{\textrm{0}}$, P$_{\textrm{1}}$,
P$_{\textrm{2}}$, ...
\item For x in Sat, input x to P$_{\textrm{0}}$
\item Test %
\footnote{If it does not give a correct answer.%
} : input x to P$_{\textrm{1}}$
\item Test : input x to P$_{\textrm{2}}$, ..., input x to P$_{\textrm{k}}$. 
\item Stop if P$_{\textrm{k}}$ gives a correct answer. %
\footnote{Outputs a satisfying line for x.%
} %
\footnote{Version of an algorithm due to \cite{B-DH}%
} %
\footnote{This isn't an algorithm because it uses Sat as an input. (Why?) Devise
an algorithm where step 2. above is substituted by : x is the binarily-coded
version of a Boolean expression in CNF. You will have to devise a
bound R' that is related to R. %
}
\end{enumerate}
Let R(k) be a function that gives a bound to that computation. %
\footnote{What is R : roughly, how far do we have to go up to a P$_{\textrm{k}}$so
that for all y instances up to x, P$_{\textrm{k}}$ correctly answers
for every such y.%
} We can define the following function :

\begin{itemize}
\item for every k, F(k) is the smallest z so that P$_{\textrm{k}}$ fails
to correctly guess a satisfying line for z. %
\footnote{The first instance, if any, where P$_{\textrm{k}}$fails.%
} %
\footnote{F(k) is the 'counterexample function', see G. Kreisel's ``no--counterexample
interpretation'' in \cite{SHO}.%
} %
\footnote{It comes up in several guises.%
} 
\item Get an enumeration of all Turing machines. 
\item Put F'(k) = 0, iff M$_{\textrm{k}}$ is not a poly machine.
\item F'(k) = F(k) iff M$_{\textrm{k}}$ is a poly machine.
\end{itemize}
As a bit of exercise, fill in the gaps : 

\begin{lyxlist}{00.00.0000}
\item [A)]Prove that F' isn't recursive. %
\footnote{Easy.%
}
\item [C)]Prove that F' grows as the Busy Beaver Function.%
\footnote{Somewhat more difficult.%
}
\item [B)]Get a condition X consistent with ZFC so that : 
\end{lyxlist}
\begin{enumerate}
\item ZFC + Condition X proves {[}F is total{]} <---> {[}F' is total{]}.
\item ZFC proves {[}F is total{]} ---> {[}Condition X{]}.
\item ZFC neither proves {[}Condition X{]} nor its negation. %
\footnote{At hand.%
}
\end{enumerate}
\begin{lyxlist}{00.00.0000}
\item [C)]Show that a function R which is primitive-recursive related to
the counter example function F above gives a bound to the \cite{B-DH}
algorithm. %
\footnote{Don't forget to claim 7.000.000 \$ instead.%
} 
\end{lyxlist}

\section{Conclusion}

Collect your prizes and honors and kudos and then go home to do some
useful work... at last !

\end{document}